\documentclass[10pt]{article}
\usepackage{graphicx}
\usepackage{lineno}
\def\Title#1{\begin{center} {\Large #1 } \end{center}}
\def\Author#1{\begin{center}{ \sc #1} \end{center}}
\def\Address#1{\begin{center}{ \it #1} \end{center}}

\newcommand\pubblock{\rightline{\begin{tabular}{l} Proceedings of the Second Annual LHCP\\ \pubnumber\\
         \pubdate  \end{tabular}}}

\newenvironment{Abstract}{\begin{quotation} \begin{center} 
             \large ABSTRACT \end{center}\bigskip 
      \begin{center}\begin{large}}{\end{large}\end{center} \end{quotation}}

\newenvironment{Presented}{\begin{quotation} \begin{center} 
             PRESENTED AT\end{center}\bigskip 
      \begin{center}\begin{large}}{\end{large}\end{center} \end{quotation}}

\def\beq{\begin{equation}}
\def\eeq#1{\label{#1}\end{equation}}
\def\eeqn{\end{equation}}

\def\beqa{\begin{eqnarray}}
\def\eeqa#1{\label{#1}\end{eqnarray}}
\def\eeqan{\end{eqnarray}}

\let\bar=\overbar

\def\Dslash{\not{\hbox{\kern-4pt $D$}}}
\def\dslash{\not{\hbox{\kern-2pt $\del$}}}

\def\msb{{\bar{\ssstyle M \kern -1pt S}}}

\usepackage{color}

\newcommand\pubnumber{ ATL-PHYS-PROC-2014-024 }

\newcommand\pubdate{\today}

\def\affiliation{
On behalf of the ATLAS Collaboration, \\
Department of Physics \\
Waseda University, Tokyo, Japan }

\begin{document}
\large
\begin{titlepage}
\pubblock

\vfill
\Title{The ATLAS Tau Trigger Performance during LHC Run 1 \\ \vspace{2mm} and Prospects for Run 2}
\vfill

\Author{ YUKI SAKURAI }
\Address{\affiliation}
\vfill
\begin{Abstract}

  Triggering on hadronic tau decays is essential for a wide variety of analyses of interesting physics processes at ATLAS. 
  The ATLAS tau trigger combines information from the tracking detectors and calorimeters to identify the signature of hadronically decaying tau leptons. 
  In Run 2 operation expected to start in 2015, the trigger strategies will become more important than ever before. 
  In this paper, the tau trigger performance during Run 1 is summarized and also an overview of the developments of Run 2 tau trigger strategy is presented.

\end{Abstract}
\vfill

\begin{Presented}
The Second Annual Conference\\
 on Large Hadron Collider Physics \\
Columbia University, New York, U.S.A \\ 
June 2-7, 2014
\end{Presented}
\vfill
\end{titlepage}
\def\thefootnote{\fnsymbol{footnote}}
\setcounter{footnote}{0}
%

\normalsize 


\section{Introduction}

The tau lepton, a lepton of third-generation with a mass of $1.78\,$GeV,
plays an important role in both precise measurement of Standard Model physics 
and search for physics beyond the Standard Model (e.g., SUSY, $W^{'}/Z^{'}$).
Especially, an efficient tau identification is crucial for measurements of the Higgs-fermion coupling constant (i.e., $H\rightarrow\tau\tau$ \cite{ref:Htautau}).

The tau lepton has a short life time ($\sim 2.9\cdot10^{-13}s$) and thus a short decay length ($c\tau = 87 \,\mu m$).
Therefore, it decays inside the beam pipe and has to be identified by its decay products.
Decay modes of the tau lepton can be classified into ``leptonic'' and ``hadronic'' decays.
The decay products of leptonic decays are one electron or muon and two neutrinos with a total branching 35\%.
Leptonic tau decays are covered by electron and muon triggers in the ATLAS trigger system \cite{ref:atlas}.
Hadronic tau decays contain hadrons and a tau neutrino in the final state with branching ratio of 65\%. 
The tau trigger aims to trigger the events containing hadronically decaying taus ($\tau_{had}$).
Because of the branching ratio, efficient triggering on $\tau_{had}$ and its improvement directly enhances the sensitivity
not only for $H\rightarrow\tau\tau$ analysis but for any other physics processes including $\tau_{had}$.

The main background of $\tau_{had}$ is jets from quarks or gluons. The QCD hadronization process produces a number of hadrons that can fake a $\tau_{had}$ signature. 
Thus, the primary goal is to achieve strong rejection power against jets in the trigger level.
QCD jets typically have a large number of tracks, while $\tau_{had}$ has only one (1-prong) or three (multi-prong) tracks from
charged pions (or kaons). Therefore, the following two characteristics of a hadronic tau decay can be used:

\vspace{-1mm}
\begin{enumerate}
  \setlength{\parskip}{0cm} 
  \setlength{\itemsep}{0cm}
\item 1 or 3 charged tracks in the core cone;
\item no tracks and energy deposition in the isolation cone,
\end{enumerate}
\vspace{-1mm}
where the core (isolation) cone is defined as the region within $\Delta R=\sqrt{\Delta\eta + \Delta\phi} < 0.2 \ (0.4)$ from the highest momentum track of the $\tau_{had}$ candidate. 

\vspace{-2mm}
\begin{figure}[htbp]
  \begin{minipage}{0.5\hsize}
    \begin{center}
      \vspace{0.6cm}
    \includegraphics[height=2cm,width=4.5cm]{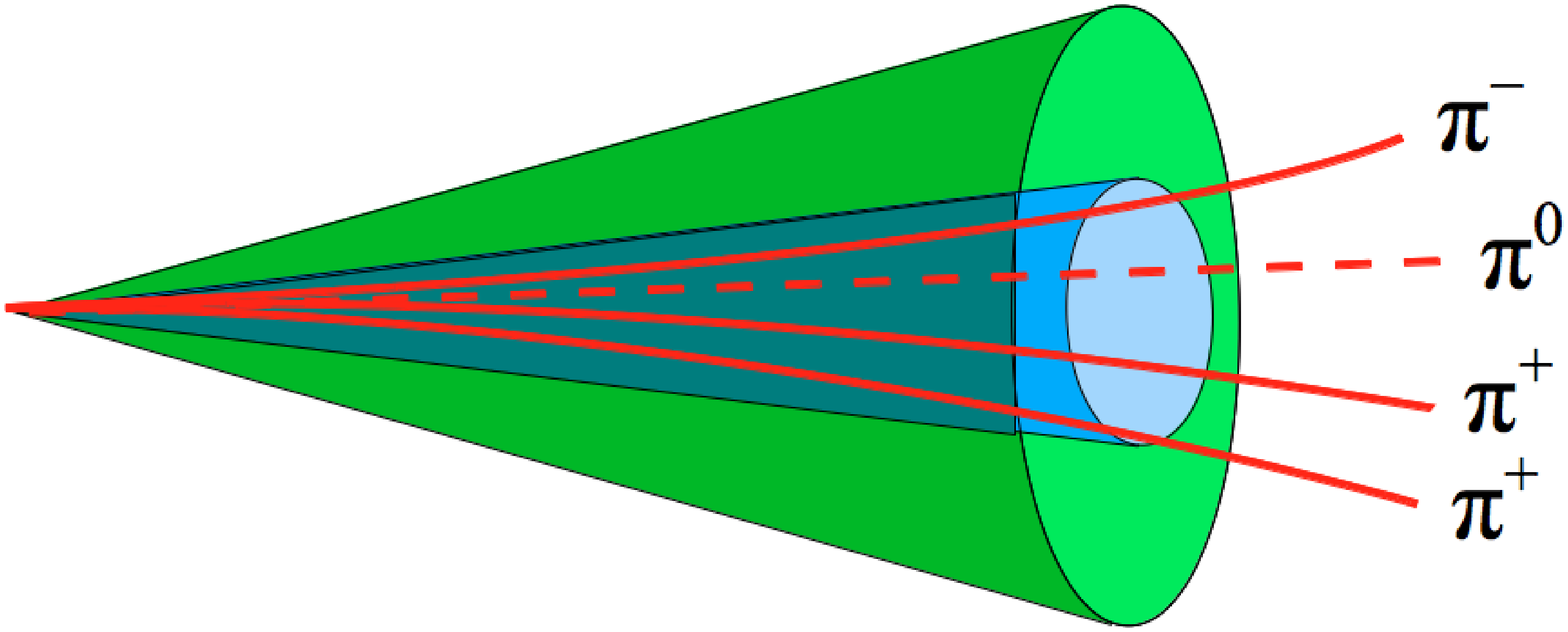}
    \end{center}
  \end{minipage}
  \begin{minipage}{0.5\hsize}
    \begin{center}
      \vspace{0.6cm}
      \includegraphics[height=2.3cm,width=4.5cm]{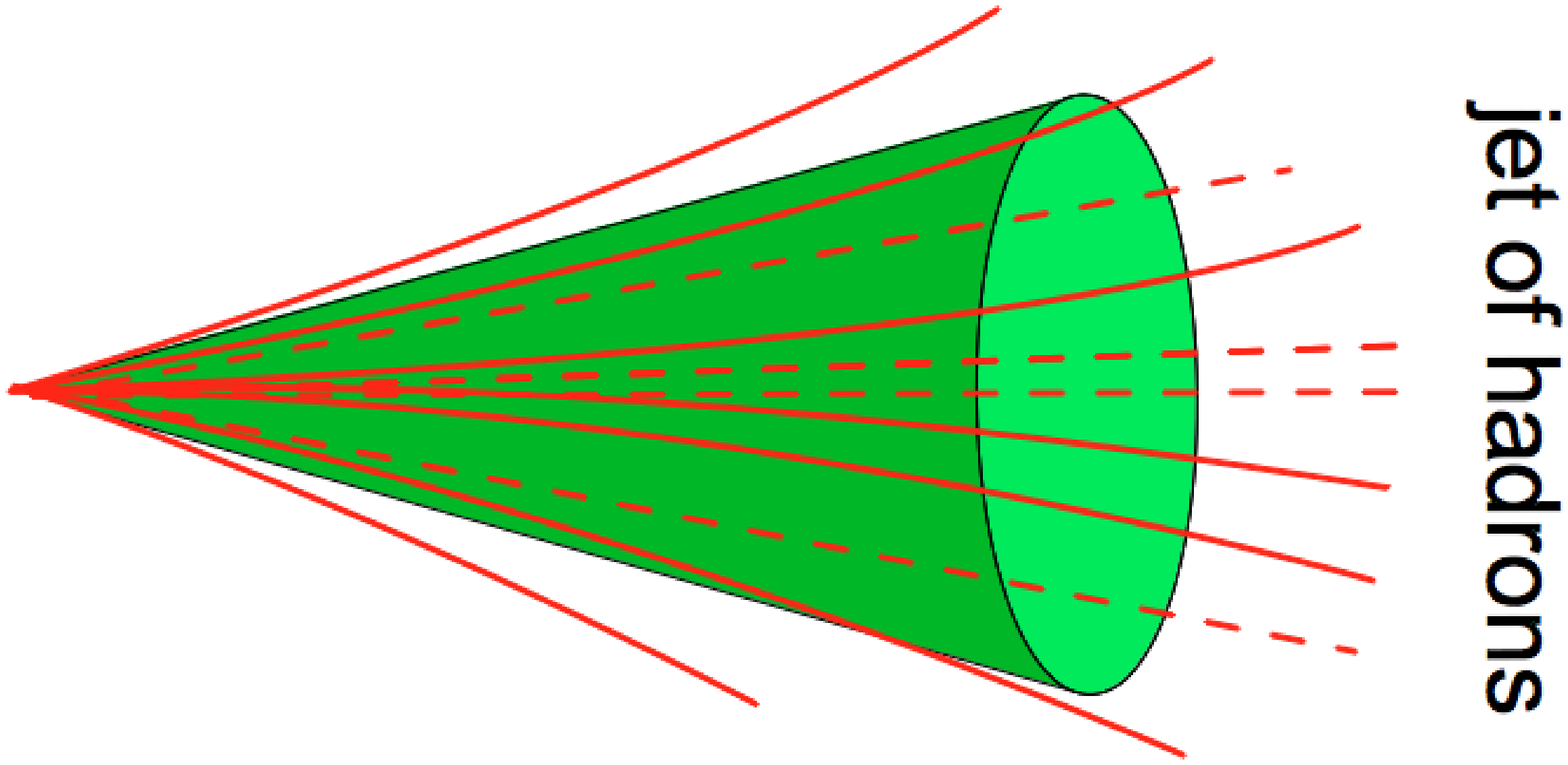}
    \end{center}
  \end{minipage}
  \label{fig:cone}   
  \vspace{-2mm}
  \caption{ Typical signature of a $\tau_{had}$ (left) and QCD jet (right) }
  
\end{figure}
\vspace{-2mm}

\section{Tau Trigger Structure in Run 1}

The ATLAS trigger system must reduce the output event rate to about 400$\,fg$Hz from an initial LHC bunch crossing rate of 20$\,$MHz (50$\,$ns bunch spacing).
To record the interesting events for physics with feasible event rate, the trigger system has three levels: Level 1, Level 2 and Event Filter \cite{ref:2011}.

The Level 1 (L1) tau trigger is a simple and fast hardware trigger that uses the information of electromagnetic (EM) and hadronic (HAD) calorimeter towers 
of approximate size of $\Delta\eta \times \Delta \phi = 0.1 \times 0.1$.
A given threshold set a requirement on the energy of $\tau_{had}$ candidate  
in $2 \times 1$ or $1 \times 2$ EM calorimeter towers and $2 \times 2$ HAD calorimeter towers.
An optional isolation requirement on the energies in the $4 \times 4$ ring outside the core region of $2 \times 2$ EM calorimeter towers can be applied to further reduce the event rate.
The cone centered at the position of $\tau_{had}$ candidate which satisfies the L1 requirements is called region of interest (RoI) and passed to the next trigger level.

The level 2 (L2) tau trigger is a software based trigger that uses the energy and track information within each RoI given by the L1 trigger.
To distinguish $\tau_{had}$ from QCD jets, several discriminating variables, such as the track multiplicity, 
the average distance between tracks and the shape of the particle showers, are used.

The Event Filter (EF) is a software based trigger using the full detector information.
A Boosted Decision Tree (BDT), a powerful, commonly used multivariate classifier, is employed to efficiently identify $\tau_{had}$.
Several discriminating variables are input to the BDT. They are synchronized with the offline identification 
as much as possible to achieve optimal signal efficiency and background reduction.

\section{Performance in Run 1}

The tau trigger was operated with different $p_{T}$ thresholds during Run 1 and their performance was studied using real data.
The trigger efficiency, defined as probability of a reconstructed and identified $\tau_{had}$ candidate at offline level to pass the tau trigger requirement, 
was measured by the tag-and-probe method in the $Z\rightarrow\tau\tau\rightarrow(\mu\nu_{\mu}\bar{\nu_{\tau}})(\tau_{had}\nu_{\tau})$ decay.
In this method, the isolated muon from the leptonic tau decay is required to tag the events, while the $\tau_{had}$ can be used to determine the tau trigger performance.
The dominant background events ({\it W}+jets, QCD) are reduced by a requirement that the transverse mass\footnote{ $m_{T} = \sqrt{ 2p_{T}^{\tau_{had}} \cdot E_{T}^{miss} \cdot (1 - cos \Delta\phi (\tau_{had},E_{T}^{miss})) }$, where $E_{T}^{miss}$ is the missing transverse energy }$\,$($m_{T}$) is greater than 50$\,$GeV and the invariant mass for a muon and a $\tau_{had}$ is in the range of 40$\,$GeV and 80$\,$GeV.

\begin{figure}[htbp]
  \begin{minipage}{0.5\hsize}
    \begin{center}
      \includegraphics[width=8cm]{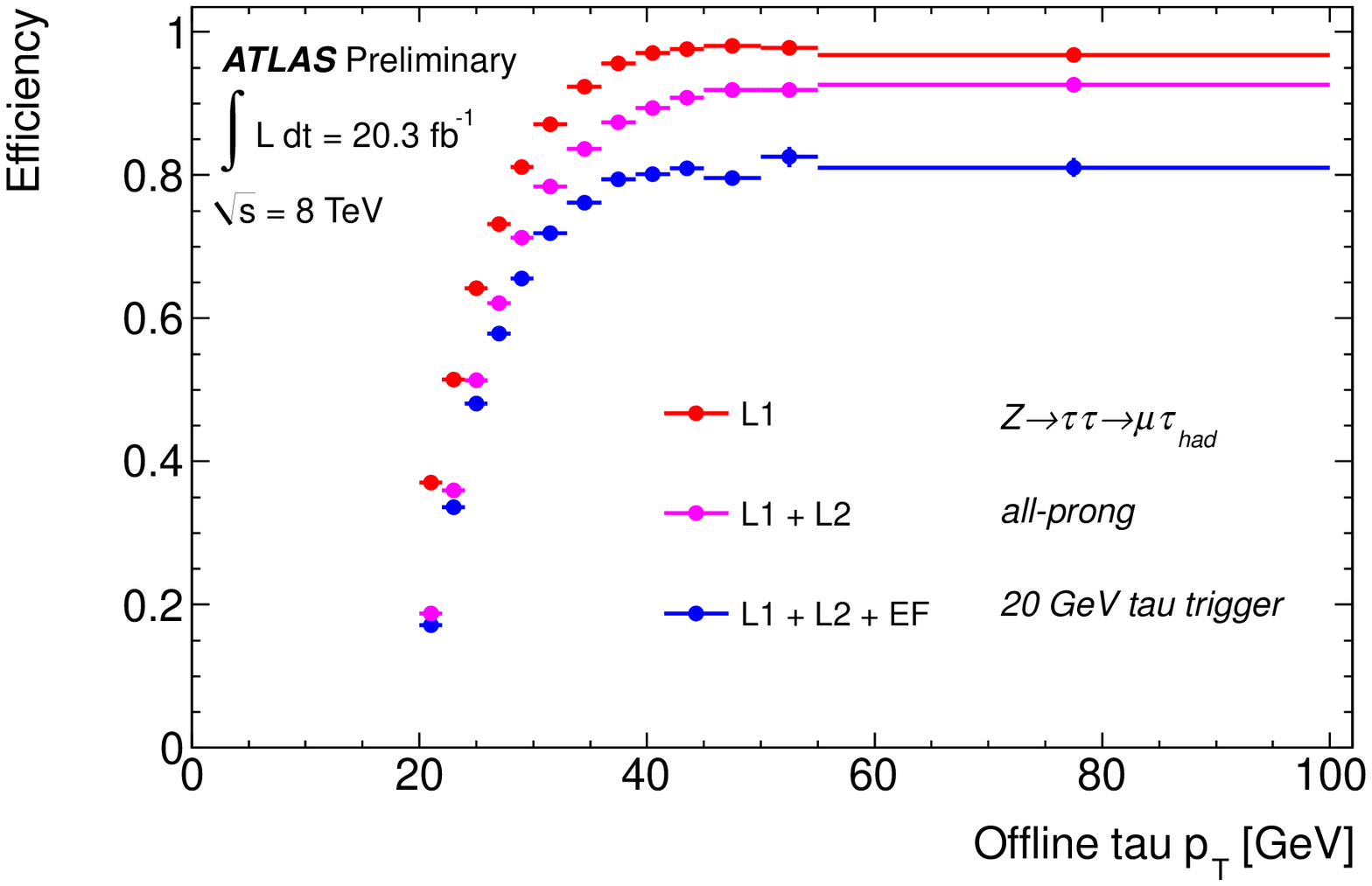}
    \end{center}
  \end{minipage}
  \begin{minipage}{0.5\hsize}
    \begin{center}
    \includegraphics[width=8cm]{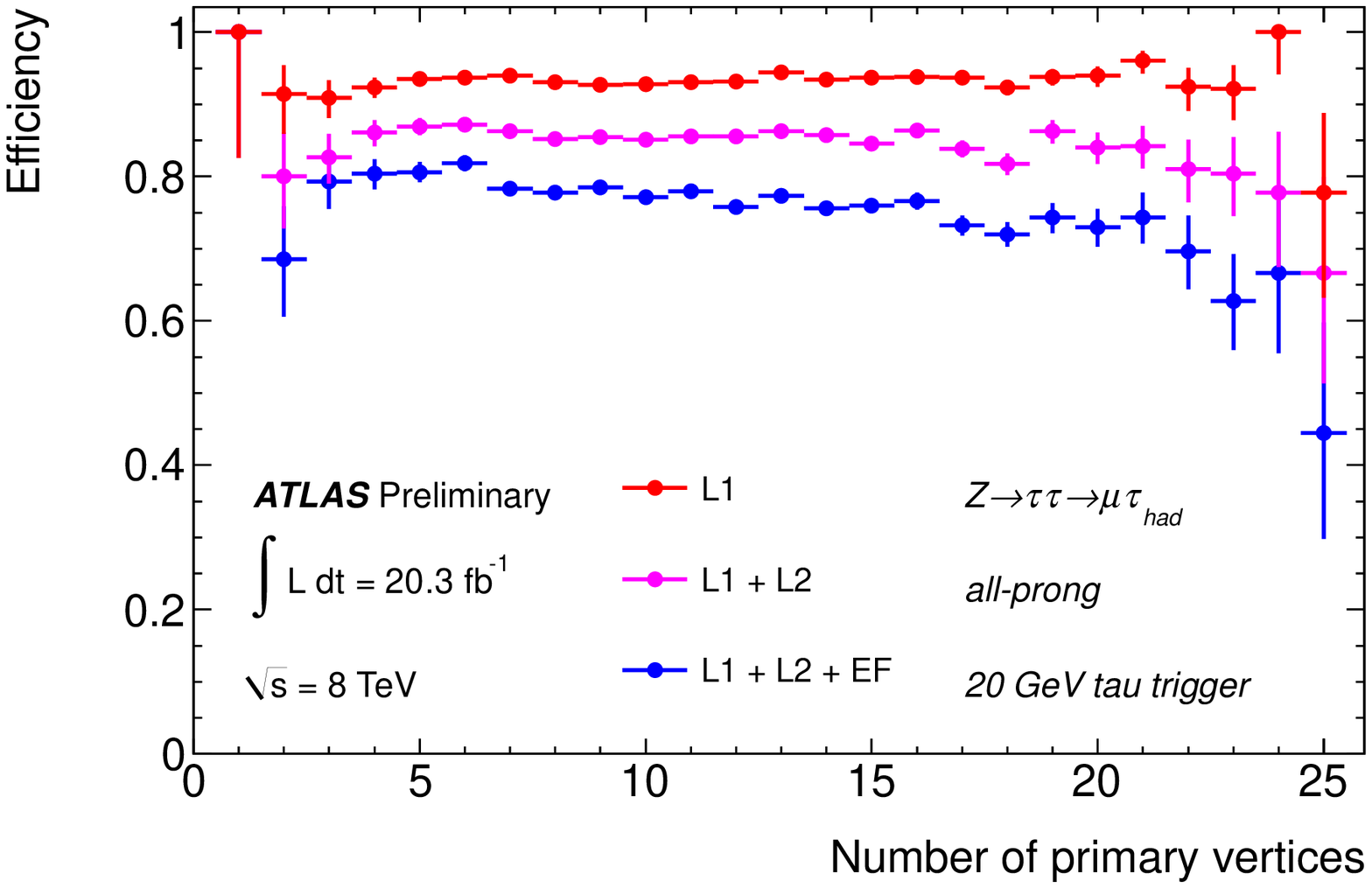}
    \end{center}
  \end{minipage}
  \caption{  The tau trigger efficiencies of $p_{T}$ threshold of 11$\,$GeV at L1, and 20$\,$GeV at L2 and EF as functions of the $p_{T}$ of offline $\tau_{had}$ (left) and number of primary vertices (right). }
  \label{fig:eff}
\end{figure}

The tau trigger efficiencies of the $p_{T}$ threshold of 11$\,$GeV at L1, and 20$\,$GeV at L2 and EF are shown in Fig.\ref{fig:eff} \cite{ref:twiki}.
There is no significant loss in high pile-up events.
Therefore, the Run 1 tau trigger succeeded to keep the signal efficiency, even in the high pile-up environment.

\section{Prospect for Run 2}

With the higher bunch crossing rate (25$\,$ns bunch spacing) and peak luminosity ($2 \times 10^{34} cm^{-2}s^{-1}$),
the trigger strategies in Run 2 will become more important than in Run 1 for stable data taking.
Having such a higher event rate, one of the major challenges will be to keep the event rate at the designed level without increasing the thresholds significantly.
Therefore, several improvements have been developed:

\vspace{-1mm}
\begin{itemize}
\item {\bf Topological Trigger at Level 1 (L1Topo)}
\end{itemize}
\vspace{-1mm}

To cope with the higher event rate at L1,
the new Topological Processor called L1Topo will be installed in the L1 trigger system for Run 2.
The L1Topo will provide an additional rate reduction using topological information.
Particularly for the di-tau (electron+tau) trigger, angular information between $\tau_{had}$-$\tau_{had}$ (e-$\tau_{had}$) will be used in the L1Topo.

\vspace{-1mm}
\begin{itemize}
\item {\bf Merging of L2 and EF}
\end{itemize}
\vspace{-1mm}

Thanks to availability of a number of new CPU's for Run 2, L2 and EF trigger systems are migrated to a single trigger system called High Level Trigger (HLT).
It makes more the flexible trigger system and possible to bring online algorithms closer to offline algorithms.

\vspace{-1mm}
\begin{itemize}
\item {\bf Topological clustering algorithm (TopoClustering)}
\end{itemize}
\vspace{-1mm}

\begin{figure}[htbp]
  \begin{minipage}{0.5\hsize}
    \begin{center}
      \includegraphics[width=7cm]{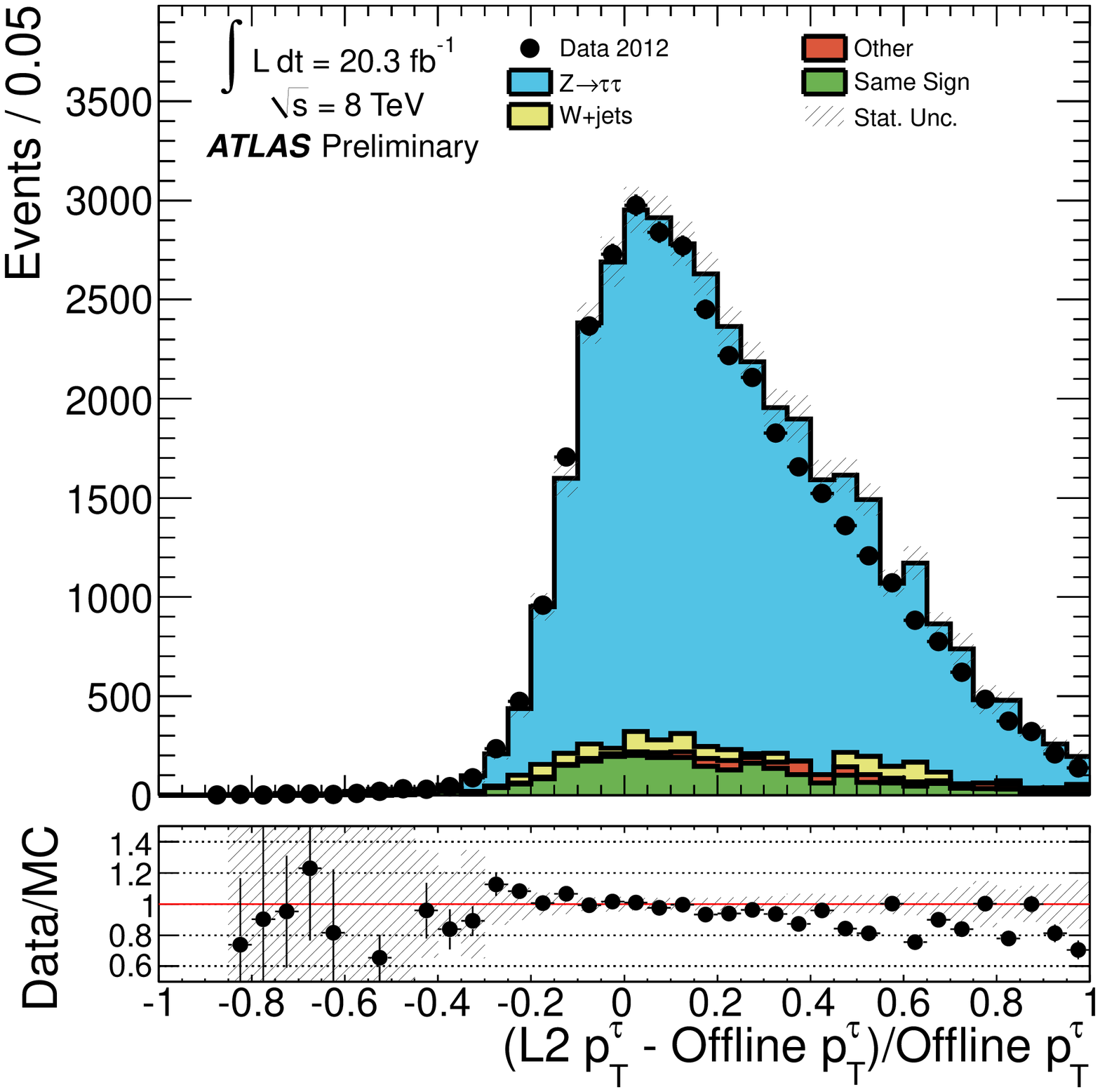}
    \end{center}
  \end{minipage}
  \begin{minipage}{0.5\hsize}
    \begin{center}
    \includegraphics[width=7cm]{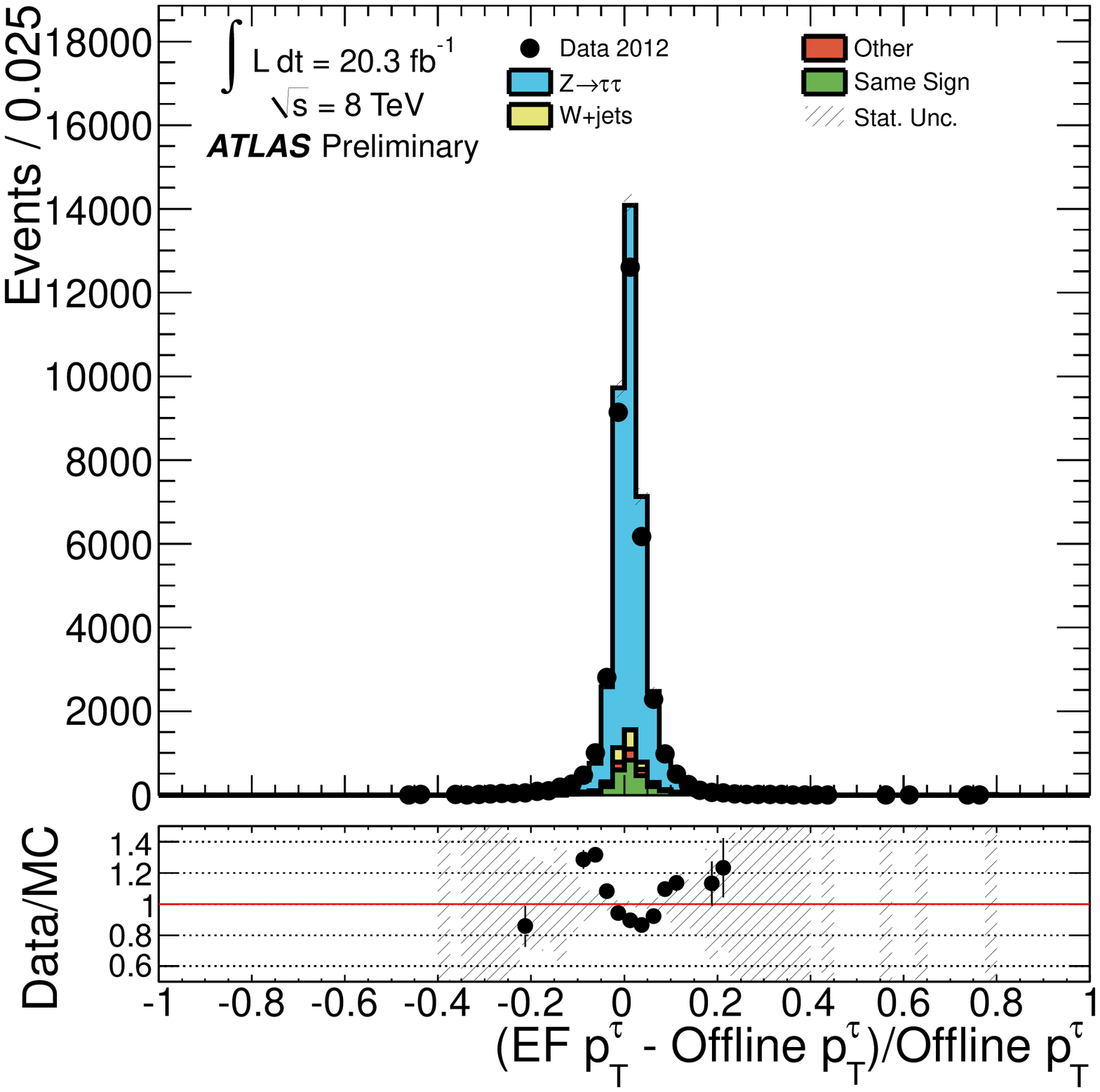}
    \end{center}
  \end{minipage}
  \caption{ $p_{T}$ resolution of L2 (left) and EF (right) with respect to the offline $p_{T}$ at Run 1. 
    The L2 energy is reconstructed by summing the energies of the calorimeter tower, 
    while the EF energy reconstruction uses TopoClustering method. }
  \label{fig:res} 
\end{figure}

In Run 1, the energy reconstruction method at L2 is the sum of the energies in calorimeter tower,
while EF uses a topological clustering algorithm called TopoClustering \cite{ref:topoclustering} with better resolution than L2 (Fig. \ref{fig:res}).
The energy reconstruction in Run 2 will be improved by applying TopoClustering algorithm from the beginning of HLT.

\begin{itemize}
\item {\bf FastTracker (FTK)}
\end{itemize}

The FTK \cite{ref:ftk} is one of the upgrade project and it will give full track information at the beginning of HLT.
The FTK will become available in barrel ($\eta<1.1$) region from 2016, and in full region from 2018.
The tau trigger performance will be improved by applying selection using FTK tracks and 
pile-up corrections based on vertices reconstructed with FTK tracks.

\vspace{3mm}

\section{Conclusion}

In this paper, the tau trigger performance during Run 1 and an overview of the developments of Run 2 tau trigger strategy are summarized.
The Run 1 tau trigger succeeded to keep the signal efficiency even in the high pile-up environment, as shown in Fig. \ref{fig:eff}.
Thanks to this, the Run 1 tau trigger was used by several analyses including $\tau_{had}$ in final state (e.g. $H\rightarrow\tau\tau$).
In order to cope with higher event rate in Run 2, several improvements have been developed, they are showing good signs for the stable data taking.

\end{document}